\magnification\magstep1
\tolerance 1600
\parskip=6pt
\centerline{\bf The Beacon of Kac-Moody Symmetry for Physics}
\vskip30pt
\centerline { L. Dolan }
\vskip 12 pt
\centerline{\it Department of Physics and Astronomy,
University of North Carolina}
\centerline{\it Chapel Hill, North Carolina 27599-3255, USA}
\vskip30pt

Although the importance of developing a mathematics which transcends
practical use was already understood by the Greeks over 2000 years ago, 
it is heartening even today when  
mathematical ideas created for their abstract interest 
are found to be useful in formulating descriptions of nature.  
Historically, the idea of symmetry has its scientific origin in 
the Greeks' discovery of the five regular solids, which
are remarkably symmetrical.
In the 19th century, this property was codified in the mathematical
concept of a {\it group} invented by Galois and then that of 
a {\it continuous group} by Sophus Lie. 

In 1967, Victor Kac (MIT) then working in Moscow [K] and Bob Moody 
(Alberta) [M] 
independently enlarged the paradigm of classical Lie algebras, 
resulting in new algebras which are infinite-dimensional.
The representation theory of a subclass of the algebras, the affine 
Kac-Moody algebras, has 
developed into a mature mathematics.

By the 1980's, these algebras had been taken up by physicists 
working in the areas of 
elementary particle theory, gravity, and two-dimensional phase transitions 
as an obvious framework from which to consider  
descriptions of non-perturbative solutions of gauge theory, vertex emission
operators in string theory on compactified space, integrability in 
two-dimensional quantum field theory, and conformal field theory. 
Recently Kac-Moody algebras have been shown to serve as duality symmetries 
of non-perturbative strings appearing
to relate all superstrings to a single theory. The infinite-dimensional Lie
algebras and groups have been suggested as candidates for a unified symmetry
of superstring theory. 

In addition to this wide application to physical theories, the Kac-Moody
algebras are relevant to number theory and modular forms. They occur in the
relation between the sporadic simple Monster group and the symmetries of 
codes, lattices, and conformal field theories [Gr,CN]. Much in the same way
that certain finite groups were understood to be the symmetry groups of the
regular solids, the Monster and certain affine Lie algebras are seen to be 
automorphisms of conformal field theories [FLM,DGM].  
Kac's generalization of Weyl's character formula for the affine algebras 
leads to a deeper understanding of Macdonald combinatoric identities 
relating infinite products and sums [Mac,K1]. 
Topological properties of the groups corresponding to the affine algebras 
have been analyzed [Ge,PS], as well as varieties related to the
the singularity theory of Kac-Moody algebras [K2,Na,Sl].
Several reviews emphasize the physical applications of the algebras [D,GO,J].

\eject
{\bf Familiar Symmetry} 

In physics, the symmetries of a theory 
provide significant information
about the general solution of the system. 
For example, a general solution to Laplace's equation $\nabla^2
U(x,y,z)=0$ can be compounded from particular solutions in a 
coordinate system suggested by the symmetry of the problem.
For a spherical conductor, limited to the case where $U$ is 
$U(r,\theta)$, the solution can be expressed in terms of specific
functions of $r$ and the Legendre polynomials $P_n(\theta)$.
In quantum field theories, useful applications of symmetry 
include the invariance imposed on the 
regularization of an approximation scheme, 
symmetry transformations among different specific solutions, 
and Noether constants of the motion used to label them.
As evidenced in the works of Weyl [Wey], van der Waerden [Wa], 
and Wigner [Wi], finite groups and then continous Lie groups with  
their associated algebras were central to the understanding of atomic 
spectra and the formulation of quantum mechanics in the 1930's
[Wigh]. 

Familiar symmetries
reflect not only spacetime invariance such as the Poincar\'e group or the
general linear group of gravity, but also internal gauge invariance. 
For example, gauge invariance given by the group $U(1)$ results in  
the conservation of electric charge in Maxwell's equations.
The notions of connection and curvature in modern differential geometry
describe the gravity, electromagnetic, and Yang-Mills gauge theories 
of the fundamental interactions of physics.
Already in the 1920's Kaluza and Klein [KK] suggested a 
unification of the classical theories of gravity and electromagnetism 
by observing that the Maxwell field (later identified as a connection 
on a principal U(1) fiber bundle) 
could be understood as extra components of gravity defined in five spacetime
dimensions, with the fifth dimension made compact. 
This related the symmetries of gravity and electromagnetism 
in four dimensions and
provided one of Einstein's models for a unified field theory. 
The Kaluza-Klein mechanism for curling up dimensions is eventually seen
to accomodate even higher-dimensional models such as string theory. 

In the 1970's, high energy physicists pursued Lie algebra theory
as a valuable tool to characterize all the gauge interactions. 
These are now 
understood to be $SU(3)$ for the strong force (which describes
the interacations between quarks, which
are the constituents of hadrons such as the proton), 
and $SU(2)\times U(1)$ for both the weak and electromagnetic 
interactions of quarks 
and leptons (such as the electron).  
This is an important 
feature of the standard model of
particle physics [YM,We,Sa,Gl].
{\it Grand unification} was an effort to combine these symmetries as
subgroups of a unifying group such as $SU(5)$.
{\it Superstring unification} provides an alternative mechanism to 
combine symmetries.

The interpretation of fundamental force laws in terms of group theory is
now commonly understood in terms of E. Noether's theorem which
identifies the elements of the Lie algebra with the charges 
conserved in the interactions [No].
The symmetry is used to label the physical states: the eigenvalues of 
the Cartan subalgebra are the quantum numbers of the elementary particles.
The quantum numbers are known as charge, spin, hypercharge, isospin, etc.
depending on which group is being considered.  
Modern high energy theorists effectively think of the elementary 
particles of the strong, weak and electromagnetic forces
as irreducible representations of the direct product of the Poincar\'e 
group and $SU(3)\times SU(2)\times U(1)$. 

Selection rules derived from the conserved charges
limit allowed transitions 
of quantum numbers. Explicit solutions of four-dimensional 
quantum field theory transition amplitudes, however,
are known only in a perturbative
expansion, i.e. they are not known exactly. 
This weak coupling perturbative approximation is 
extremely useful in electro-weak theory, but the strong coupling 
problem of why the quarks are confined inside the hadrons
remains more elusive. 

In two spacetime dimensions the situation is different.
Both discrete systems (for e.g. the Ising model) and continuous theories
(for eg. the sine-Gordon equation and the principal chiral models) 
can be solved exactly, i.e. they are
integrable due to the occurrence of an infinite number of 
conservation laws. Essentially each theory has as many constants of 
motion as it has degrees of freedom. For example, the affine algebra has
been used to construct a general class of solutions for the 
Korteweg-deVries equation and to linearize the periodic Toda lattice. 
In the case of a finite parameter algebra, the method of orbits of Lie
groups has led to the quantization of the integrable Toda chains [Ko]. 
These stunning results of mathematical physics
suggested that infinite-dimensional algebras as well as the finite-parameter 
symmetry algebras might be important for physical theories. 
\vskip10pt
{\bf Kac-Moody Symmetry and Conformal Field Theory}

The theory for the finite-dimensional semisimple 
Lie algebras was worked out by E. Cartan and Killing about
a hundred years ago. They associated with each algebra 
a finite integer matrix with positive-definite conditions.
Generalizing the Cartan matrix by relaxing the positivity conditions, 
one can obtain the Kac-Moody algebras, of which a subclass constitutes the
conventional Lie algebras.
Another subclass of Kac-Moody algebras referred to as {\it affine}
replaces positive-definite with positive-semidefinite; and the corresponding
structure and representation theory has a precise analogy to that of the
semisimple finite-dimensional algebras. The commutation relations for the
affine algebra are:
$$[T_n^a, T_m^b] = i f_{abc} T_{n+m}^c + k n\delta_{n,-m}\delta_{ab}$$
where $n,m\in{\bf Z}$, $T_0^a\in g$ for any semisimple Lie algebra $g$
with structure constants $f_{abc}$, so $1<a,b,c<{\rm dim} g$. 
The central extension $k$ is proportional 
to the identity and is related to the level of the affine algebra. 
Since the generators are moments of currents, i.e. densities
$T^a(z) =\sum_n T_n^a z^{-n-1}$, their commutators can 
be expressed equivalently in terms of an operator product expansion
for $|z| > |\zeta|$ by
$$T^a(z) T^b(\zeta) = (z-\zeta)^{-1} i f_{abc} T^c(\zeta) + (z-\zeta)^{-2}
k \delta_{ab} + {\rm regular\hskip3pt terms}\,.$$
This expansion is reminiscent of Gell-Mann's current algebra approach
for local particle symmetries. This emphasized that not only do 
the {\it charge} operators $Q^a$ form an algebra, such as $SU(3)$ in the
eightfold way used to classify elementary particles, but also the 
local {\it currents} $J^a(x)$ satisfy algebraic commutation
relations [AD,TJG]. 

Infinite-dimensional 
Kac-Moody algebras which are not affine include those
called {\it hyperbolic},
and recent
connections with superstring theory have led to a better understanding of
their representations [GNW] as well.  

For a Kac-Moody algebra, the
Cartan matrix, and thus the Cartan subalgebra remains finite-dimensional. 
Each affine algebra $\hat g$ has a finite-dimensional semisimple Lie 
algebra
$g$ whose rank, i.e. number of simple roots, 
is one less than that of the whole affine algebra. Therefore, theories with
the larger symmetry have one additional 
quantum number called the level. 
We can also extend the affine algebra to include an element called the
derivation operator $-L_0$ where $[L_0, T_n^a] = -n T_n^a$, whose  
eigenvalue can be thought of as the affine charge, 
and measures the mass level in physical theories. The derivation
is the zero mode of another infinite-dimensional algebra, the Virasoro
algebra. These generators can be derived via the Sugawara construction and 
are bilinear in the affine generators
$L(z)\equiv 1/(2k + c_\psi)\, \sum_a  {\scriptstyle{\times \atop \times}}
T^a (z) T^a (z) {\scriptstyle{\times \atop \times}}$, so that
$$\eqalignno{[L_n, L_m] &= (n-m) L_{n+m} + {c\over 12} (n^3-n)\delta_{n,-m}\cr
[L_n, T_m^a ]&= -m T_{n+m}^a\cr}$$
where the central charge $c$ is related to the level of the affine
algebra and $f_{abc} f_{abe} = c_\psi \delta_{ce}$. 
The Virasoro algebra corresponds to  
the infinite-dimensional conformal symmetry of two-dimensional conformal
field theories. These field theories are relevant to two-dimensional
spin systems at a phase transition and to the two-dimensional
world sheet traced out by a string as it moves through time [BPZ].
In statistical models, representations of the conformal symmetry
lead to an identification of the highest weights with critical exponents
[FQS]. It is when physicists began to ask detailed questions about the
representation theory of the infinite-dimensional algebras, that progress
took off in those physical theories. 

The quantum number associated with the affine charge 
is the conformal weight. 
An irreducible representation of an affine algebra $\hat g$,
with non-zero level, is labelled by
the level and the highest weight, and
is comprised of an infinite tower of irreducible
representations of $g$. 
The tower is a stack of different `grades',
with a grade being distinguished by its conformal weight. 
The pattern is worked out for the first few grades in terms of 
$su(2)$ representations in fig. 1.
The $g$ content of a given representation can be expressed by
multiplicity formulas [KP]. For example, 
the multiplicity $\phi_\kappa (x)$ for the level one highest weight
singlet representation of $\hat su(2)$ is 
$$\phi_\kappa (x) = \sum_{s=0}^\infty n_\kappa (s) \, x^s =
x^{\kappa^2} (1-x^{2\kappa +1}) \prod_{\ell = 1}^\infty (1-x^\ell)^{-1}\,;
\quad \kappa =0,1,2,\ldots\,,$$
where 
$n_\kappa (s)$                 
is equal to the number of $su(2)$ multiplets with highest weight
$\kappa$ in grade s.
These algebras and their representations are applicable to  
physical theories with an infinite number of states, such as string theory
or bound states (strong coupling limits) of gauge theories. See fig. 1. 

The `value added' in moving from Lie groups to Kac-Moody algebras in 
physical theories is that
larger symmetry groups give more information about the solution.
In the case of conformally invariant systems for example, 
there are primary fields which transform as
$$[L_n, \phi (z) ] = z^{n+1}{d\phi(z)\over dz} + h (n+1) z^n\phi(z)$$
and amplitudes can be computed exactly as follows.
Since the vacuum state $|0\rangle$ satisfies $ L_n |0\rangle = 0\,,\,
n\ge -1$, then $L_0\phi(0) |0\rangle = h \phi(0) |0\rangle $,
$L_n\phi(0) |0\rangle = 0\,, n> 0$ and the two 
point function is determined up to a multiplicative constant,
$$\langle 0| \phi (z_1) \phi (z_2) |0\rangle =
r^{-h} e^{-i\theta h} \langle 0| \phi (1) \phi (0) |0\rangle$$
where $z_1 - z_2 = r e^{i\theta}$.
\vskip 10pt
{\bf Integrable Systems and Gauge Theories}

The first hint that strong interaction gauge theory might contain a hidden 
symmetry came from an observation by Polyakov [P] that a functional 
formulation of the non-abelian theory was similar to the local equations
of the two-dimensional integrable chiral models. In this approach, the 
fundamental role of
the gauge field $A_\mu^a (x)$ is replaced by the path
dependent field $\psi[\xi] = P exp(\oint A d\xi) =
P exp\{\oint ds \dot\xi_\mu (x) A_\mu(\xi(s))\}$, 
an element of the 
holonomy group, i.e. a path dependent element of the gauge group. 
Here $A_\mu\equiv A_\mu^a T^a$ where $T^a$ are the elements of $su(N)$.
A functional differential equation for $\psi [\xi]$ can be derived:
$${\delta\over{\delta\xi_\mu(s)}} (\psi^{-1}
{\delta\psi\over{\delta\xi_\mu(s)}}) = 0$$
when $A_\mu$ satisfies the Yang-Mills equations of motion 
$D_\mu F_{\mu\nu} =0$.
These loop equations look like the chiral model equations:
$$\partial_\alpha (g^{-1}\partial_\alpha g) = 0\,,$$
which have an infinite number of symmetries [LP]; and the algebra of 
a set of 
these symmetries was shown to be given by that of the positive modes of 
an affine Kac-Moody algebra [D]. 
It was then natural to conjecture that the same symmetry algebra should 
be responsible for such transformations in both theories,
and that representations of Kac-Moody
algebras will give information about the gauge theory bound state spectrum,
which is comprised of confined sets of gluons. 
This is consistent with the 
belief that bound states of gluons lie on linear trajectories,
dictating an infinite number of them with 
a linear relation between the mass and the 
spin of the states,
which is a generic feature of a string theory.
Certainly knowledge of the complete symmetry 
group of the theory is important information.

Properties of instantons in two-dimensions and in  
the four-dimensional self-dual Yang-Mills equations  
demonstrate similarity in the symmetry of the models as well [A].
Affine symmetry has also been identified as potentially useful 
in describing the physical particle content 
of extended supergravity models whose scalar fields parameterize
a symmetric space $G/H$, where $G$ is a non-compact global symmetry group
with $H$ as its maximal compact subgroup [EGGZ]. 

Recently, an approach using discrete duality symmetries in supersymmetric
gauge theories has led to some analytic control over a mechanism
for describing quark-gluon
confinement [SW]. These theories, when viewed as the low energy limits of
string theories, lead the way to discovering the larger string dualities 
and the fundamental symmetry group of the string.

\vskip 10pt
{\bf String Theory}

In 1968, around the time Kac and Moody formulated their algebras,  
Veneziano [V] wrote down particle scattering amplitudes
with certain analyticity properties. 
The amplitudes are functions of the momenta of the particles,
and can be derived as the scattering of
modes of a massless relativistic string, a one-dimensional 
object whose length
is characterized by the nature of the coupling. For closed strings, which 
contain gravity as well as the quarks and leptons, 
the scale is the Planck length, $10^{-33}$ centimeters, and
the different modes of oscillation describe the different 
particles, much in the same way that different modes on a violin
string result in different notes. 
Later as the interacting string
picture, which set up the unitarity and general consistency of 
quantum string theory was being developed [Man], 
mathematicians constructed irreducible
representations of the affine algebras. They used
string theory vertex operators whose momenta were restricted to
take on discrete values, such as the points on the root lattice of $g$. 
The discretization of momenta implies a compactified, closed, periodic 
condition on the conjugate position space. See fig. 2. 
The number of compactified dimensions is equal to the rank of $g$. 

In the early 1980's it was
understood by physicists that since the superstring 
was quantizable only in ten dimensions, discrete momenta 
corresponding to 
the curling up of six of these was the obvious Kaluza-Klein 
answer to achieve a string theory relevant for physics in 
four spacetime dimensions. See fig. 3.

The level one representations were constructed from the 
Veneziano open bosonic string Fock space oscillators, with the 
compactification size $R$ fixed to be the string scale $\alpha '$
[FK,Se]. See fig. 4.
Earlier, representations had been found for
twisted affine algebras with use of the Corrigan-Fairlie [CF] oscillators 
corresponding to strings with
different boundary conditions, i.e. one end of the string is held
at a fixed position [LW]. 
Higher level quark model representations were given in terms of fermionic 
oscillators [BH].

These realizations were then incorporated into realistic string models,
which appear to have a chance of describing standard model physics.
The heterotic string [GHMR] makes use of level one affine  
representations, while compactifications
of the Green Schwarz superstring [GS] employ representations with
level equal to the dual Coxeter number of $g$ [BDG]. 
In both models, the zero mode generators which are elements of $g$
correspond to the Yang-Mills gauge symmetry of the particle spectrum.

At the time of the rebirth of string theory in 1985 [GSW],
physicists had been
working hard on solving the problem of quark confinement in four-dimensional
non-abelian gauge theory.
It was believed that the added structure in string theory,
whose low energy limit was gauge theory, might prove sufficient to 
formulate a non-perturbative solution. In the last year or so,
developments concerning duality symmetries have  
occurred which carry this program further. 

Duality symmetry that maps between strong and weak coupling in gauge 
theories and in string theory has been tested [Sen]. 
The simplest notion of this kind of symmetry surfaced originally 
in the study of integrable models,
where solvability and strong-weak maps were seen to go hand in hand.
In fact, the Kramers-Wannier self-duality of spin models [KW], such as
the Ising model and the X-Z model can be used to construct an 
infinite commuting algebra of conserved charges [DG]. A kind of 
Kramers-Wannier duality was incorporated by Montonen and Olive [MO]
to formulate a conjecture for electric-magnetic duality in gauge theories,
since the Dirac quantization condition for monopoles fixes the electric
and magnetic couplings to be inversely proportional. 
This duality occurs naturally in a supersymmetric gauge
theory and has been extended to  
the strong and weak coupling limits of
string theory. The duality transformations can be used to relate 
different superstring models to one another [Du,HT,W], so that
in fact there may be just one structure whose quantum ground state 
describing elementary particles is unique. 

It now seems feasible [Sc,GNW]
that the infinite discrete set of duality symmetries of string theory 
may be related to both $E_9$, i.e. the affine
$E_8$ algebra, and to a hyperbolic Kac-Moody Lie algebra $E_{10}$.
The idea of using Kac-Moody symmetry to get at  
non-perturbative information in theories of particle physics 
remains viable.
\vskip 10pt
{\bf Physics and Mathematics}

Many of the properties of the Kac-Moody algebras were 
rediscovered in physical theories.
They supply the precise Fock space states
and field operators which provide a practical  
way to compute structure constants and construct representations 
and realizations. 
The occurrence of these symmetries in physics has
been useful to the development of the mathematical theory.
Nonetheless, the sophisticated mathematical structure which has evolved
serves as a beacon to physicists in that it signals that the theories
under investigation probably have further properties which are not yet 
obvious but may be instrumental in understanding the consistency of nature.  
\vskip 10pt
\noindent{\bf References}

\item{[AD]} S. Adler and R. Dashen, {\it Current algebras and applications to
particle physics}, Benjamin, New York, 1986.
\item{[A]} M.F. Atiyah, {\it Instantons in two and four dimensions},
Comm.Math.Phys. {\bf 93} (1984) 437-451.
\item{[BH]} K. Bardakci and M.B. Halpern, {\it New dual quark models},
Phys.Rev.{\bf D3} (1971) 2493-2506.
\item{[BPZ]} A. Belavin, A. Polyakov, and A. Zamolodchikov,
{\it Infinite conformal symmetry in two-dimensional quantum field theory},
Nucl.Phys.{\bf B241} (1984) 333-380. 
\item{[BDG]} R. Bluhm, L. Dolan and P. Goddard, {\it A new method of 
incorporating symmetry in superstring theory}, Nucl.Phys.{\bf B289} 
(1987) 364-384.
\item{[CN]} J.H. Conway and S.P. Norton, {\it Monstrous moonshine},
Bull.Lond.Math.Soc.{\bf 11} (1979) 308-339. 
\item{[CF]} E. Corrigan and D. Fairlie,
{\it Off-shell states in dual resonance theory},
Nucl.Phys.{\bf B91} (1975) 527-545. 
\item{[D]} L. Dolan, 
{\it Kac-Moody algebra is hidden symmetry of chiral models},
Phys.Rev.Lett.{\bf 47} (1981) 1371-1374;
{\it Kac-Moody algebras and exact solvability in hadronic physics},
Phys.Rep.{\bf 109} (1984) 1-94. 
\item{[DGM]} L. Dolan, P. Goddard and P. Montague,
{\it Conformal field theory and twisted vertex operators},
Nucl.Phys.{\bf B338} (1990) 529-601; 
{\it Conformal field theory, triality and the Monster group},
Phys. Lett. {\bf B236} (1990) 165-172.
\item{[DG]} L. Dolan and M. Grady, {\it Conserved charges from
self-duality}, Phys.Rev.{\bf D25} (1982) 1587-1604.
\item{[Du]} M.J. Duff, {\it Duality rotations in string theory},
Nucl.Phys.{\bf B335} (1990) 610-620.
\item{[EGGZ]} J. Ellis, M.K. Gaillard, M. Gunaydin and B. Zumino,
{\it Supersymmetry and non-compact groups in supergravity}, 
Nucl.Phys.{\bf B224} (1983) 427.
\item{[FK]} I. Frenkel and V.G. Kac, {\it Basic representations of 
affine lie algebras and dual models}, Inv.Math.{\bf 62} (1980) 23-66.
\item{[FLM]} I. Frenkel, J. Lepowsky and A. Meurman, {\it Vertex 
operator algebras and the Monster}, Academic Press, New York, 1988.
\item{[FQS]} D. Friedan, Z. Qui and S. Shenker,  
{\it Conformal invariance and critical exponents in two dimensions},
Phys.Rev.Lett.{\bf 52} (1984) 1575-1578.   
\item{[GNW]} R. Gebert, H. Nicolai and P. West, {\it Multistring vertices
and hyperbolic Kac-Moody algebras}, hep-th/9505106.
\item{[Ge]} R. Geroch, 
{\it A method for generalizing new solutions of Einstein's 
equations I\&II}, J.Math.Phys.{\bf 12} (1971) 918-924; 
J.Math.Phys.{\bf 13} (1972) 394-404.
\item{[Gl]} S. Glashow, 
{\it Towards a unified theory},
Rev.Mod.Phys.{\bf 52} (1980) 539-543. 
\item{[GO]} P. Goddard and D. Olive, {\it Kac-Moody and Virasoro algebras
in relation to quantum physics}, Int.Jour.Mod.Phys.{\bf A1} (1986) 303-414.
\item{[GS]} M.B. Green and J.H. Schwarz, {\it Supersymmetrical string 
theories}, Phys.Lett.{\bf B109} (1982) 444-448.
\item{[GSW]} M.B. Green, J.H. Schwarz and E. Witten, 
{\it Superstring theory, Vol. 1\&2}, Cambridge University Press, 1988.
\item{[Gr]} R.L. Griess, {\it The friendly giant}, Inv.Math.{\bf 69}
(1982) 1-102. 
\item{[GHMR]} D. Gross, J. Harvey, E. Martinec, and R. Rohm,
{\it The free heterotic string}, Nucl.Phys.{\bf B256} (1985) 253-284.
\item{[HT]} C. Hull and P. Townsend, {\it Unity of superstring dualities},
hep-th/9410167; {\it Enhanced gauge symmetries in superstring theories},
hep-th/9505073.
\item{[J]} B. Julia, {\it Supergeometry and Kac-Moody algebras} in
Vertex Operators in Mathematics and Physics, MSRI publication \# 3,
Springer, Heidelberg, 1984, 393-410.
\item{[K]} V.G. Kac, {\it Simple graded algebras of finite growth}, 
Func.Anal.Appl.{\bf 1} (1967) 328-329.
\item{[K1]} V.G. Kac, {\it Infinite-dimensional Lie algebras},
Third edition, Cambridge University Press, 1990.
\item{[K2]} V.G. Kac, {\it Infinite root systems, representations 
of graphs and invariant theory}, Inv.Math.{\bf 56} (1980) 57-92.
\item{[KP]} V.G. Kac and D.H. Peterson, {\it Infinite-dimensional Lie 
algebras, theta funtions and modular forms},
Adv.Math.{\bf 53} (1984) 125-264.
\item{[KK]} Th. Kaluza, {\it On the problem of unity in physics},
Sitzungsber.Preuss.Akad.Wiss.Berlin. Math. Phys.{\bf K1} (1921) 966-;
O. Klein, {\it Quantum theory and 5-dimensional theory of relativity},
A.Phys.{\bf37} (1926) 895-.
\item{[Ko]} B. Kostant, {\it On Whittaker vectors and representation
theory}, Inv.Math.{\bf48} (1978) 101-184.
\item{[KW]} H.A. Kramers and C.H. Wannier, 
{\it Statistics of the two-dimensional ferromagnet},
Phys.Rev.{\bf 60} (1941) 252-276.
\item{[LW]} J. Lepowsky and G. Wilson, 
{\it Construction of the affine Lie algebra $A_1^{(1)}$},
Com.Math. Phys.{\bf 62} (1978) 43-53.
\item{[LP]} M. L\"uscher and K. Pohlmeyer, {\it Scattering of massless
lumps and non-local charges in the two-dimensional classical $\sigma$-
model}, Nucl.Phys.{\bf B137} (1978) 46-54.
\item{[Mac]} I.G. Macdonald, {\it Affine Lie algebras and modular forms},
Seminare Bourbaki Exp. 577, Lecture Notes in Mathematics Vol. 901, 258-65.
Springer-Verlag, New York, 1981.
\item{[Man]} S. Mandelstam, 
{\it Interacting-string picture of dual resonance models},
Nucl.Phys.{\bf B64} (1973) 205-235.
\item{[MO]} C. Montonen and D. Olive, 
{\it Magnetic monopoles as gauge particles?},
Phys.Lett.{\bf B72} (1977) 117-120; D. Olive, 
{\it Exact electromagnetic duality}, hep-th/9508089. 
\item{[M]} R.V. Moody, {\it Lie Algebras associated with generalized Cartan
matrices}, Bull.Am.Math. Soc. {\bf 73} (1967) 217-221.  
\item{[Na]} H. Nakajima, {\it Instantons on ALE spaces, quiver varieties, and
Kac-Moody algebras}, Duke Mathematical Journal {\bf 76}(2) (1994), 365-416.
\item{[No]} E. Noether, {\it Invariant variational problems}, 
Transport Theory and Statistical Physics{\bf 1}(3) (1971), 183-207;
M.A. Tavel, tr. This is a translation of Emmy Noether's original paper 
in Nachr. d. K\"onig, Gesellsch. d. Wiss. zu G\"ottingen, Math-phys. 
Klasse (1918) 235-257. 
\item{[P]} A. Polyakov, {\it Gauge fields as rings of glue}, 
Nucl.Phys.{\bf B164} (1980) 171-188.
\item{[PS]} A. Pressley and G. Segal, {\it Loop Groups}, Clarendon Press,
Oxford, 1986.
\item{[Sa]} A. Salam, 
{\it Gauge unification of fundamental forces},
Rev.Mod.Phys.{\bf 52} (1980) 525-538.
\item{[Sc]} J.H. Schwarz, {\it Classical symmetries of two-dimensional models},
hep-th/9503078, hep-th/9506076; {\it String theory symmetries} hep-th/9503127. 
\item{[Se]} G. Segal, {\it Unitary representations of some infinite-
dimensional groups}, Comm.Math. Phys.{\bf 80} (1981) 301-342.
\item{[SW]} N. Seiberg and E. Witten, {\it Monopoles, duality and
chiral symmetry breaking in $N=2$ supersymmetric QCD}, 
Nucl.Phys.{\bf B431} (1994) 484-550, hep-th/9408099.
\item{[Sen]} A. Sen, {{\it Dyon-monopole bound states, self-dual harmonic
forms on the multi-monopole moduli space, and SL(2,Z) invariance in
string theory}, Phys.Lett.{\bf B329} (1994) 217-221. 
\item{[Sl]} P. Slodowy, {\it Simple singularities and simple algebraic groups},
Lecture Notes in Mathematics Vol. 815,
Springer-Verlag, New York, 1980.
\item{[TJG]} S. Treiman, R. Jackiw and D. Gross, {\it Lectures on current
algebra and its applications}, Princeton University Press, 1972.
\item{[V]} G. Veneziano, {\it Construction of a crossing-symmetric
Regge-behaved amplitude for linearly rising trajectories},
Nuovo Cimento{\bf 57A} (1968) 190-197.
\item{[Wa]} B.L. Van der Waerden, {Group theory and quantum mechanics}, 
Springer-Verlag, Berlin and New York, 1974.
\item{[We]} S. Weinberg, 
{\it Unified theory of weak and electromagnetic interaction},
Rev.Mod.Phys. {\bf 52} (1980) 515-523.
\item{[Wey]} H. Weyl, {\it The theory of groups and quantum mechanics}, Dover,
New York, 1950; H.P. Robertson, tr.
\item{[Wigh]} A.S. Wightman,{\it Obituary of E. Wigner},
AMS Notices, {\bf 42} (July 1995) 769-771.
\item{[Wi]} E. Wigner, {\it Group theory and its applications to the 
quantum mechanics of atomic spectra}, Academic Press, New York, 1959.
\item{[W]} E. Witten, {\it String theory dynamics in various dimensions},
hep-th/9503124.
\item{[YM]} C.N. Yang and R.L. Mills, 
{\it Conservation of isotopic spin and isotopic gauge invariance},
Phys.Rev.{\bf 96} (1954) 191-195.

\vskip 10pt

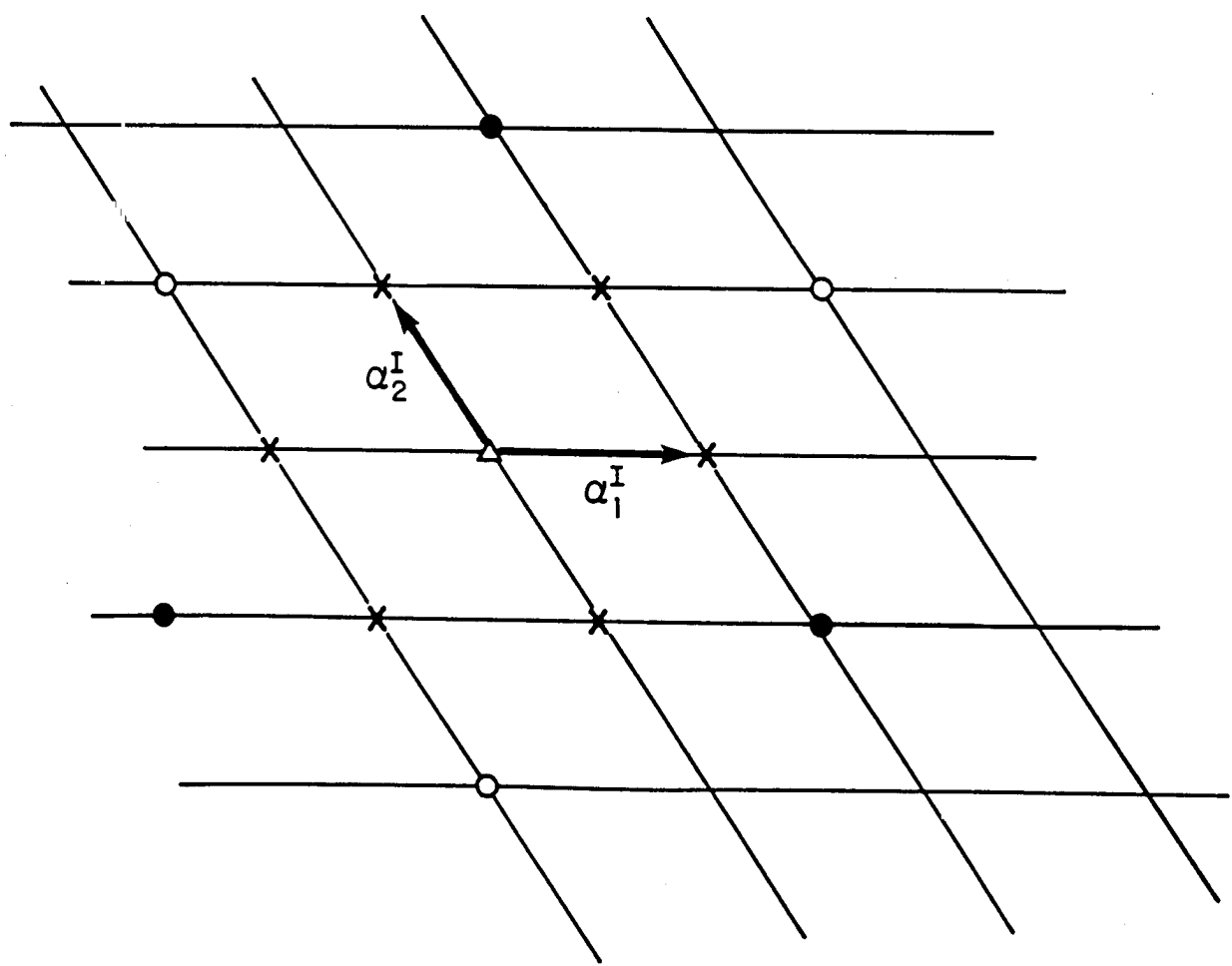
\bye